\documentclass[preprint2,numberedappendix]{emulateapj-rtx4}
\usepackage{graphicx,bm,natbib,color,url}
\usepackage{soul}
\graphicspath{{./fig/}{./png/}}

\newcommand{\Eq}[1]{Equation~(\ref{#1})}

\newcommand{\Fig}[1]{Figure~\ref{#1}}

\newcommand{\kurt}{\mbox{kurt}}
\newcommand{\Mpc}{\,{\rm Mpc}}
\newcommand{\G}{\,{\rm G}}
\newcommand{\s}{\,{\rm s}}

\newcommand{\iii}{{\rm i}}
\newcommand{\dd}{{\rm d}}
\newcommand{\EM}{E_{\rm M}}
\newcommand{\EC}{E_{\rm C}}
\newcommand{\FM}{F_{\rm M}}
\newcommand{\FC}{F_{\rm C}}
\newcommand{\BB}{\bm{B}}
\newcommand{\JJ}{\bm{J}}
\newcommand{\kk}{\bm{k}}
\newcommand{\rr}{\bm{r}}
\newcommand{\uu}{\bm{u}}
\newcommand{\nab}{\bm{\nabla}}
\newcommand{\arxiv}[2]{ #1, arXiv:#2.}
\newcommand{\sprd}[2]{ #1, PhRvD, submitted, arXiv:#2.}

\newcommand{\yprl}[3]{ #1, PhRvL, #2, #3}
\newcommand{\yprd}[3]{ #1, PhRvD, #2, #3}

\newcommand{\ysph}[3]{ #1, SoPh, #2, #3}
\newcommand{\yapj}[3]{ #1, ApJ, #2, #3}
\newcommand{\yapjl}[3]{ #1, ApJL, #2, #3}
\newcommand{\ymn}[3]{ #1, MNRAS, #2, #3}
\newcommand{\ygafd}[3]{ #1, GApFD, #2, #3}
\newcommand{\yjcap}[3]{ #1, JCAP, #2, #3}
\newcommand{\yrpp}[3]{ #1, RPPh, #2, #3}
\newcommand{\yjour}[4]{ #1, #2, #3, #4}
\newcommand{\ybook}[3]{ #1, #2 (#3)}

\begin{document}

\title{The turbulent stress spectrum in the inertial and subinertial ranges}

\author{Axel Brandenburg$^{1,2,3,4}$\thanks{E-mail: }
and Stanislav Boldyrev$^{5,6}$}

\affil{
$^1$Nordita, KTH Royal Institute of Technology and Stockholm University, Roslagstullsbacken 23, SE-10691 Stockholm, Sweden\\
$^2$Department of Astronomy, AlbaNova University Center, Stockholm University, SE-10691 Stockholm, Sweden\\
$^3$JILA and Laboratory for Atmospheric and Space Physics, University of Colorado, Boulder, CO 80303, USA\\
$^4$McWilliams Center for Cosmology \& Department of Physics, Carnegie Mellon University, Pittsburgh, PA 15213, USA\\
$^5$Department of Physics, University of Wisconsin -- Madison, 1150 University Avenue, Madison, WI 53706, USA\\
$^6$Space Science Institute, Boulder, CO 80301, USA
}

\begin{abstract}
For velocity and magnetic fields, the turbulent pressure and, more
generally, the squared fields such as the components of the turbulent 
stress tensor, play important roles in astrophysics.
For both one and three dimensions, we derive the equations relating the
energy spectra of the fields to the spectra of their squares.
We solve the resulting integrals numerically and show that for turbulent 
energy spectra of Kolmogorov type, the spectral slope of the stress 
spectrum is also of Kolmogorov type.
For shallower turbulence spectra, the slope of the stress spectrum
quickly approaches that of white noise, regardless of how blue
the spectrum of the field is.
For fully helical fields, the stress spectrum is elevated
by about a factor of two in the subinertial range, while that
in the inertial range remains unchanged.
We discuss possible implications for understanding the spectrum of 
primordial gravitational waves from causally generated
magnetic fields during cosmological phase transitions
in the early universe.
We also discuss potential diagnostic applications to the interstellar medium,
where polarization and scintillation measurements characterize the
square of the magnetic field.
\end{abstract}


\maketitle

\section{Introduction}

In aeroacoustics, the stress tensor of the turbulent velocity field plays an
important role in sound generation.
Its theory goes back to the work of \cite{Lig52a,Lig52b},
whose equation is also used in astrophysics to describe
the heating of stellar coronae by pressure waves
excited in the outer convection zones of stars \citep{Ste67}.
Similarly, in the early universe, the velocity stress
and also the combined stress of velocity and
magnetic fields can be responsible for driving
primordial gravitational waves \citep{KKT94,DFK00}.
In that case, it is important to relate spectra of the turbulence
to the spectra of the kinetic and magnetic stresses in order to compute
the spectrum of the gravitational waves \citep{Gogo07,RPMBKK19}.

Empirically, it was known that a velocity or magnetic field with
a Kolmogorov-type power law spectrum produces a similar
spectrum for the stress, except
that in the subinertial range, where the spectral energy increases
with wavenumber $k$, the spectral slope of the stress never
increases with $k$ faster than
for white noise \citep{RPMBKK19}, even if the turbulence has a
blue spectrum.
This has important implications for understanding the gravitational wave
production at very low frequencies from primordial magnetic fields.
Such magnetic fields can be generated at the electroweak phase
transition \citep[see][for a review]{Sub16}, but their spectrum
would be steeper than that of white noise \citep{DC03}
and could not readily explain the shallower
white noise spectrum of the stress.

There are different conventions for expressing energy spectra.
In this paper, we always present the energy per uniform
(linear as opposed to logarithmic) wavenumber interval, so the
mean energy density is therefore $\int_0^\infty E(k)\,\dd k$.
In three dimensions, a white noise spectrum is then proportional to $k^2$.
At some wavenumber $k_*$, the spectral energy begins to decline again.
The value of $k_*$ determines the scale where most of the energy resides.
At an even higher wavenumber $k_{\rm D}$, dissipation becomes important
and the spectral energy falls off exponentially.
The spectral range from $k_*$ to $k_{\rm D}$ is called the inertial range.
Its spectral slope is determined by the nature of turbulence.
For Kolmogorov turbulence, it would be proportional to $k^{-5/3}$.
The spectral range below $k_*$ is called the subinertial range.
Here, the flow tends to be completely uncorrelated,
and this is what determines its spectral slope.

In the early universe, when it was just $10^{-11}\s$ old,
magnetic fields are believed to have been produced with a blue 
subinertial range spectrum proportional to $k^4$ \citep{DC03}.
This is because the magnetic field is divergence free,
so the magnetic field itself does not have a white noise spectrum,
but it must be the magnetic vector potential that does.
Since the magnetic field is the curl of the vector potential,
the spectrum of magnetic energy has an extra $k^2$ factor
as compared to white noise,
which is the reason why the magnetic energy spectrum is
steeper than that of white noise.

There are other applications where the knowledge of the
spectrum of a squared function is important.
An example is the magnetic pressure, which can lead to a modulation
of the gas pressure and the gas density in the interstellar medium
and hence to interstellar scintillation \citep{LG01}.
Similarly, the square of the magnetic field perpendicular to the
line of sight affects dust polarization as well as synchrotron
radiation.
Both dust and synchrotron emission, as well as interstellar
scintillation can provide useful turbulence diagnostics in
astrophysics, provided we understand the relationship between
the spectra of the magnetic field and its square.

The purpose of the present paper is to derive the relationship
between the spectrum of the turbulence and that of the
resulting stress.
Our calculations are independent of the physical model of the
turbulence and apply equally to fluid and magnetohydrodynamic
turbulence.
With the help of several examples, we illustrate the detailed
crossover behavior between different power laws.
In all cases, we ignore the temporal evolution of the fluctuations.
The temporal correlations are important for the radiation produced
by turbulence, e.g., the gravitational waves \cite[]{Gogo07}, where
the turbulent stress tensor enters as a source in the wave equation.
Studying this in detail will be the subject of a separate investigation.
Here we focus instead on the specific relationships between
the spectra of a field and that of its stress found in the
numerical simulations of \cite{RPMBKK19}.
To illustrate the nature of the problem, it is useful to begin
with a simple example of a one-dimensional scalar field and turn
then to three-dimensional cases for scalar and vector fields,
with and without helicity.
The calculations are relatively straightforward, but we are not
aware of earlier work addressing this question.

\section{A one-dimensional example}

Let us consider the fluctuations of a scalar field (e.g., temperature,
chemical concentration, etc) $\theta(x)$ as a function of position $x$. 
We write $\theta(x)$ in terms of its Fourier transform as 
\begin{eqnarray}
\theta(x)=\int \tilde{\theta}(k)\,e^{\iii kx}\frac{\dd k}{2\pi}.
\end{eqnarray}
Due to spatial homogeneity, the correlation function
of the field can be written as
\begin{eqnarray}
\langle \tilde{\theta}(k)\tilde{\theta}^\ast (k')\rangle
=2\pi E(k)\,\delta(k-k'),
\end{eqnarray}
where $E(k)$ is the energy spectrum of $\theta$.
Its Fourier transform yields the two-point correlation function,
\begin{eqnarray}
\langle\theta(x)\theta(x')\rangle
=\int E(k)\,e^{\iii k(x-x')}\frac{\dd k}{2\pi}, 
\end{eqnarray}
and therefore
\begin{eqnarray}
\langle\theta^2(x) \rangle=\int E(k)\frac{\dd k}{2\pi}.
\end{eqnarray}

Consider now the fluctuations of the squared field 
\begin{eqnarray}
\phi(x)=\theta^2(x).
\end{eqnarray}
We are interested in the two-point correlation function
of $\phi(x)$.
We now make an important simplifying assumption (for which a physical justification will be provided later) that the four-point correlation function of $\theta$ can be split into two-point correlation functions analogously to the Gaussian rule.
We then obtain
\begin{eqnarray}
\label{phi_corr}
\langle \phi(x)\phi(x')\rangle = \langle\theta^2 \rangle^2 + 2\langle\theta(x)\theta(x') \rangle^2.
\end{eqnarray}
In order to find the energy spectrum of $\phi$, we
Fourier transform Equation~(\ref{phi_corr}) to obtain
\begin{eqnarray}
\langle \phi(x)\phi(x')\rangle
=\int F(k)e^{\iii k(x-x')}\frac{\dd k}{2\pi},
\end{eqnarray}
where
\begin{eqnarray}
F(k)=2\pi \langle\theta^2 \rangle^2\delta(k)
+2\int E(k-k')E(k')\frac{\dd k'}{2\pi}.
\end{eqnarray}
The first term could be removed by subtracting the average
of $\langle \theta^2\rangle^2$.

Let us assume we know the spectrum $E(k)$.
Our question concerns the resulting spectrum $F(k)$.
Specifically, we may think of a piece-wise power law of the form
$E(k)\propto k^\alpha$, where $\alpha$ is positive for $0<k<k_*$, and
negative for $k_*\leq k\leq k_{\rm D}$, so that the energy is contained
mostly at the scale $k_*^{-1}$, which is the outer scale of fluctuations.
We expect $F(k)$ to be asymptotically also of piecewise power law
form, $F(k)\propto k^\beta$ within a certain $k$-range.
For $k>0$ we have
\begin{equation}
F(k)=2\int_{-\infty}^\infty E(k')\,E(k-k')\, \frac{\dd k'}{2\pi},
\label{convol}
\end{equation}
where we have highlighted the fact that the integration over $k'$
goes from $-\infty$ to $+\infty$. 

At small wavenumbers $k\ll k_*$, the integral in
Equation~(\ref{convol}) is dominated by the scales $k'$
comparable to the outer scale $k_*$, so we may expand 
\begin{eqnarray}
E(k'-k)\approx E(k')-k\frac{\dd E(k')}{\dd k'} + \frac{1}{2}k^2\frac{\dd^2 E(k')}{\dd k'^2}. \end{eqnarray}
We then obtain from Equation~(\ref{convol}) the asymptotic behavior of $F(k)$ at small wavenumbers as $F(k)\approx c_1 -c_2 k^2$, where $c_1$ and $c_2$ are positive constants. This means that the spectrum $F(k)$ is flat at small $k$, that is, $\beta=0$.   

In order to find the asymptotic behavior at large wavenumbers, $k\gg k_1$,
we note that, if the energy spectrum in this interval is $E(k)\propto k^{\alpha}$,
and $-3 <\alpha < -1$, then the correlation function of $\theta$
behaves at small scales as 
\begin{eqnarray} \label{corr1}
\frac{\langle \theta(x)\theta(x')\rangle}{\langle \theta^2\rangle}
\approx 1 - \left|\frac{x-x'}{L}\right|^{-\alpha -1},
\end{eqnarray}
where $L\sim 1/k_1$ is a scale
comparable to the outer scale of the fluctuations.
The square of this correlation function then scales as
\begin{eqnarray}\label{corr2}
\frac{\langle \theta(x)\theta(x') \rangle^2}
{\langle \theta^2\rangle^2} \approx
1 - 2\left|\frac{x-x'}{L}\right|^{-\alpha -1},
\end{eqnarray}
where we have expanded the right-hand side in the small parameter
$|x-x'|/L$.
Therefore, asymptotically at large $k$, the spectrum $F(k)\sim k^\beta$
should scale with the same scaling exponent as the original energy spectrum
$E(k)$, that is, $\beta=\alpha$.

Expressions (\ref{corr1}) and (\ref{corr2}) allow us to provide a
physical motivation for splitting the fourth-order correlation functions
of $\theta$ in the pair-wise ones in formula~(\ref{phi_corr}).
For that, consider the Fourier component of the $\phi$ field,
\begin{eqnarray}\label{phik}
\phi(k)=\int \tilde{\theta}(k')\tilde{\theta}(k-k')\frac{\dd k'}{2\pi}.
\end{eqnarray}
One can ask what typical wavenumbers $k'$ and $k-k'$ contribute to this integral.
The first possibility would be to have both wavenumbers
of the same order, $k'\sim k-k'\sim k/2$.
The second possibility is to have one of these numbers much larger
than the other one, say $k' \approx k$ and $k-k'\approx 0$.
Since for the Kolmogorov spectrum, the intensity of fluctuations
declines rapidly with wavenumber, the dominant contribution
is expected to come from the second possibility.
This means that the fluctuating fields $\tilde{\theta}(k)$
contributing to the integral (\ref{phik}) have rather disparate wavenumbers.
Since the assumption of locality of Kolmogorov turbulence implies that the small-scale fluctuations are uncorrelated from the large-scale ones, we may average the $\tilde{\theta}$ fields in $\phi$ independently, which formally leads to the ``Gaussian splitting rule'' resulting in Equation~(\ref{phi_corr}).

To illustrate the resulting slope of $F(k)$ for given $E(k)$, we
consider as a first example
\begin{equation}
E(k)=\left\{\!\!
\begin{array}{ll}
&k^{-2}  \quad \mbox{for $1\leq k\leq 100$},\\
&0 \quad \quad \mbox{otherwise}.
\end{array}
\right.
\end{equation}
We compute the convolution in \Eq{convol} through
multiplication of the Fourier transformed 
quantities, i.e., through their autocorrelation
function,
\begin{equation}
\tilde{F}(x)=|\tilde{E}(x)|^2,
\end{equation}
where $\tilde{E}(x)=\int e^{\iii kx}E(k)\,\dd k/2\pi$, and likewise for $\tilde{F}(x)$.
Note that the Fourier integral is carried out from $-\infty$ to $+\infty$ and that $E(k)$ is symmetric about $k=0$.
We evaluate the integral in \Eq{convol} numerically.
The energy-carrying wavenumber in our example is $k_*\equiv1$.

\begin{figure}[t!]\begin{center}
\includegraphics[width=\columnwidth]{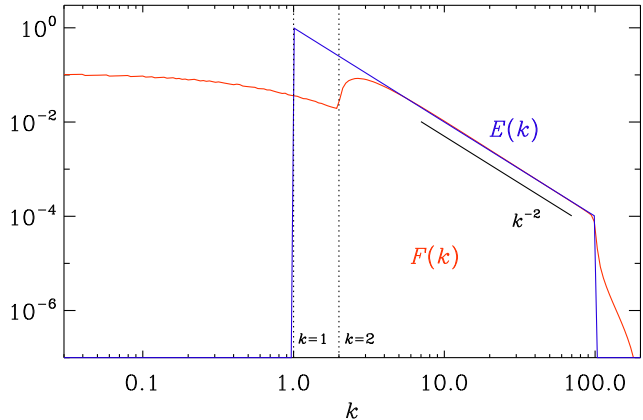}
\end{center}\caption[]{
Numerically computed $F(k)$ (red) for
$E(k)=k^{-2}$ (blue) for $1\leq k\leq 100$ (and zero otherwise).
The vertical solid and dotted lines mark $k=1$ and $2$,
respectively.
}\label{int0}\end{figure}

The result is plotted in \Fig{int0}.
We see that in the range $5<k<100$, we have $F(k)\approx E(k)$.
At $k=2$, the profile of $F(k)$ has a sharp dip, which results
in a local maximum at $k\approx3$, before approaching $E(k)$.
At small values of $k$, however, $F(k)$ always has a flat spectrum.

In our second example, we define the spectrum as
\begin{equation}
E(k)=\frac{|k|^{\alpha_1}}{[1+|k|^{(\alpha_1-\alpha_2)q}]^{1/q}}
e^{-(|k|/k_{\rm D})^2},
\label{DoublePower}
\end{equation}
where we allow for an exponential cutoff at the dissipation
wavenumber $k_{\rm D}=100$ and a power law subinertial range
of the form $k^{\alpha_1}$ with $\alpha_1=10$, $4$, $2$, $1$,
and $0$.
Here we have included the exponents $q$ and $1/q$ with $q=4$
to sharpen the cross-over between the subinertial range and the
inertial range power laws.
Furthermore, $\alpha_2=-5/3$ is the exponent chosen for the
Kolmogorov-type inertial range.

\begin{figure}[t!]\begin{center}
\includegraphics[width=\columnwidth]{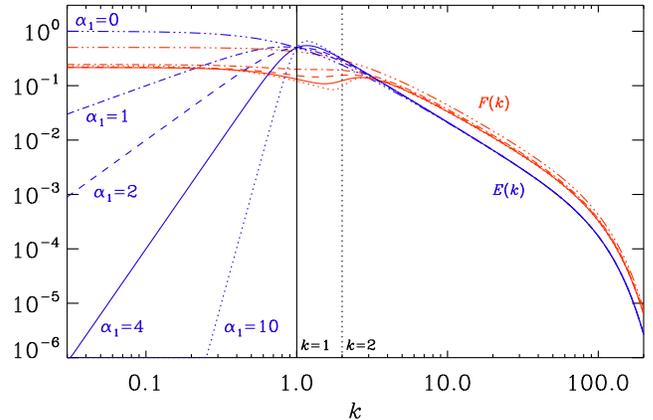}
\end{center}\caption[]{
Similar to \Fig{int0}, but for different subinertial range slopes:
$\alpha_1=0$ (triple-dot-dashed), $1$ (dot-dashed),
$2$ (dashed), $4$ (solid), and $10$ (dotted).
}\label{int}\end{figure}

The result is plotted in \Fig{int}.
In all these cases, we see that $F(k)$ has a flat spectrum
for $k\la0.5$.
We see that the dip at $k=2$ has disappeared for $\alpha_1=0$,
but it becomes stronger when $\alpha_1$ is large and
$\beta_1$ develops a white noise spectrum for small $k$.

\section{The three-dimensional case}

Next, we consider fully three-dimensional examples.
In the case of a scalar field, $\theta(\rr)$,
the derivation is similar to the one-dimensional case.
The Fourier transform of the field is defined as
\begin{eqnarray}
\theta(\rr)=\int \tilde{\theta}(\kk)e^{\iii\kk\cdot \rr}\frac{\dd^3 k}{(2\pi)^3}.
\end{eqnarray}
Then, given the correlation function of the fields
\begin{eqnarray}
\langle \tilde{\theta}(\kk)\tilde{\theta}^*(\kk')\rangle
=(2\pi)^3 I(\kk)\delta(\kk-\kk'),
\end{eqnarray}
one derives the correlation function of the quadratic field
$\phi(\rr)=\theta^2(\rr)$ in the form
\begin{eqnarray}
\langle \phi(\rr)\phi(\rr')\rangle=\int H(\kk)\frac{\dd^3 k}{(2\pi)^3},
\end{eqnarray}
where
\begin{eqnarray}
H(\kk)=(2\pi)^3\langle \theta^2 \rangle^2\delta(\kk)
+2\int \! I(\kk')I(\kk-\kk')
\frac{\dd^3 k'}{(2\pi)^3}.\quad
\end{eqnarray}

\subsection{Nonhelical vector fields}

The situation is qualitatively similar for a vector field.
Let us consider an incompressible vector field $\uu(\rr)$,
representing a velocity or magnetic field.
Its Fourier transform is defined as
\begin{eqnarray}
\uu(\rr)=\int \tilde{\uu}(\kk)\,e^{\iii\kk\cdot\rr}\frac{\dd^3k}{(2\pi)^3}.
\end{eqnarray}
We assume that the distribution of this field is homogeneous and isotropic,
so that its correlation function is given by
\begin{eqnarray}
\langle \tilde{u}^i(\kk)\tilde{u}^{*j}(\kk')\rangle
= (2\pi)^3 I(\kk)\delta(\kk-\kk')P_{ij}(\kk),
\end{eqnarray}
where we have denoted $P_{ij}(\kk)=\delta_{ij}-k_ik_j/k^2$.
The energy of this field then satisfies
\begin{eqnarray}
\langle u^2(\rr)\rangle =
\int 2I(\kk)\frac{\dd^3k}{(2\pi^3)}.
\end{eqnarray}

Similarly to the one-dimensional case, we are interested in the correlation function of the quadratic field $\phi^{ij}({\bf r})= u^i(\rr)u^j(\rr)$. Assuming that the four-point correlation functions of the $u$-field can be split into the two-point ones by using the  Gaussian rule, we get
\begin{eqnarray}\label{general}
&\langle \phi^{ij}(\kk)\phi^{*lm}(\tilde{\kk})\rangle =\nonumber\\
&\delta(\kk)\delta(\tilde{\kk})\int I(\kk')I(\kk'')P_{ij}(\kk')
P_{lm}({\bf k}'') \dd^3k' \dd^3k''+\nonumber \\
&\delta(\kk-{\tilde{\kk}})\int I(\kk-\kk')I(\kk')
P_{il}(\kk-\kk')P_{jm}(\kk')\dd^3k' + \nonumber \\
&\delta(\kk-{\tilde{\kk}})\int I(\kk-\kk')I(\kk')
P_{im}(\kk-\kk')P_{jl}(\kk')\dd^3k'. \quad
\end{eqnarray}

As an example, consider the correlation function of energy fluctuations,
\begin{equation}\label{energy_corr}
\frac{\langle\phi^{ii}(\kk)\phi^{*ll}(\tilde{\kk}) \rangle}{(2\pi)^3}
=\delta(\kk)\langle u(\rr)^2 \rangle^2+2\delta(\kk-\tilde{\kk})H(\kk),\;
\end{equation}
where 
\begin{equation}\label{H}
H(\kk)=\int I(\kk') I(\kk-\kk')
\left[1+\frac{[\kk'\cdot (\kk-\kk')]^2}{k'^2(\kk-\kk')^2}\right]\!
\frac{\dd^3 k'}{(2\pi)^3}.
\end{equation}
In a statistically isotropic case, instead of the power spectrum $I({k})$,
it is convenient to use the energy spectrum
$E({k})=4\pi k^2 I({k})$, and similarly $F(k)=4\pi k^2 H(k)$.
The above equation then becomes
\begin{equation}
F(k)=\int E(k') E(\kappa)\frac{k^2}{\kappa^2}
\left[1+\frac{(k\mu-k')^2}{\kappa^2}\right]
\frac{\dd k'\,\dd\mu}{(2\pi)^3},
\end{equation}
where
\begin{equation}
\kappa=|\kk-\kk'|=\sqrt{k^2+k'^2-2kk'\mu},
\end{equation}
and $\mu=\kk\cdot\kk'/kk'$ is the cosine of the angle between $\kk$ and $\kk'$.

\subsection{Helical vector fields}

For completeness, we also consider a more general case,
when the system is not mirror invariant.
In this case, the field correlation function has an extra term,
\begin{eqnarray}
\langle \tilde{u}^i(\kk)\tilde{u}^{*j}(\kk')\rangle
= (2\pi)^3 I(k)\delta(\kk-\kk')P_{ij}(\kk) \nonumber \\
- (2\pi)^3 J(k)\delta(\kk-\kk')\epsilon_{ijl}\iii k^l/k.
\end{eqnarray}
The last term in this expression is responsible for the helicity of the field. In particular, it enters the helicity integral
\begin{eqnarray}
\int \langle (\nab\times\uu)\cdot\uu \rangle \,\dd^3 x
= 2\int k\, J(k) \frac{\dd^3 k}{(2\pi)^3},
\end{eqnarray}
which would be zero in a mirror-invariant case.
The helicity spectral function $J(k)$ is not necessarily positive,
but it has to satisfy the realizability condition $|J(k)|\leq I(k)$.
The generalization of our results to the helical case is straightforward;
it is achieved by replacing the non-helical terms $I(k)P_{ij}({\bf k})$
in Eq.~(\ref{general}) by their helical counterparts $I(k)P_{ij}
({\bf k})-J(k)\epsilon_{ijl}\iii k^l/k$.
For example, in the correlation function of the energy fluctuations (\ref{energy_corr}) 
we will need to replace the $H({\bf k})$ function (\ref{H}) by the general expression
\begin{eqnarray}\label{H_hel}
H(\kk)=\int I(\kk') I(\kk-\kk')
\left[1+\frac{[\kk'\cdot (\kk-\kk')]^2}{k'^2(\kk-\kk')^2}\right]\!
\frac{\dd^3 k'}{(2\pi)^3} \nonumber \\
-2\int J(\kk') J(\kk-\kk')
\left[\frac{\kk'\cdot (\kk-\kk')}{k'|\kk-\kk'|}\right]\!
\frac{\dd^3 k'}{(2\pi)^3}.\quad\quad
\end{eqnarray}
Again, in a statistically isotropic case, instead of the power spectrum $J({k})$,
it is convenient to introduce the helicity spectrum defined as 
$G({k})=4\pi k^2 J({k})$.
The above equation then becomes
\begin{eqnarray}
F(k)&=&\int E(k') E(\kappa)\,\frac{k^2}{\kappa^2}
\left[1+\frac{(k\mu-k')^2}{\kappa^2}\right]\!
\frac{\dd k'\,\dd\mu}{(2\pi)^3}\nonumber \\
&&-2\int G(k') G(\kappa)\,\frac{k^2}{\kappa^2}\,
\frac{k\mu-k'}{\kappa}\,
\frac{\dd k'\,\dd\mu}{(2\pi)^3}.
\label{HelicalCase}
\end{eqnarray}
which can readily be evaluated using numerical integration.

\begin{figure}[t!]\begin{center}
\includegraphics[width=\columnwidth]{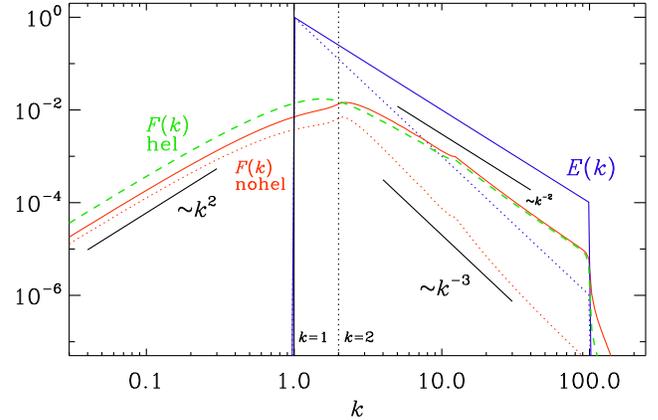}
\end{center}\caption[]{
Similar to \Fig{int0}, but for the 3D integral
for power laws $k^\alpha$ in $1<k<100$ with
$\alpha=-2$ and $-3$.
For $\alpha=-2$, $F(k)$ is also plotted
for the helical case (green dashed line).
}\label{int_term2}\end{figure}

\subsection{Examples}

In Figure~\ref{int_term2}, we show the results for the case
of a single power low, as in Equation~(12).
We consider two values for the slope $\alpha$
($-2$ and $-3$) in the range $1\leq k\leq100$.
We see that, for $k>2$, we obtain for $F(k)$ a power law,
$F(k)\propto k^\beta$, with $\beta=\alpha$,
as in the one-dimensional case.
In the range $1<k<2$, $F(k)$ is still increasing with $k$,
but the slope is slightly less steep than two.
We emphasize that this behavior is different
from that in the one-dimensional case, where we
saw instead a marked dip in $F(k)$.

In Figure~\ref{int_term2}, we also plot $F(k)$ for the case
of a fully helical field using \Eq{HelicalCase},
where $G(k)=E(k)$ is assumed.
We see that the basic features of $F(k)$ are rather similar
to the case without helicity, but there is now slightly more
power in the subinertial range, where $F(k)$ appears to be
elevated by about a factor of two.
In the inertial range, on the other hand, $F(k)$ is not
affected by the presence of helicity.

Next, we consider a spectrum with two different slopes,
$\alpha_1$  and $\alpha_2$, along with an exponential
cutoff, just as in Equation~(\ref{DoublePower}).
Again, we denote the corresponding slopes of $F(k)$ as
$\beta_1$ and $\beta_2$, respectively.
Here, we always assume a Kolmogorov inertial range spectrum
for $E(k)$, i.e., $\alpha_2=-5/3$, and we vary $\alpha_1$
from $0$ to $10$.
Physically relevant are the Saffman ($\alpha_1=2$) and
Batchelor ($\alpha_1=4$) asymptotic scalings for $k\to 0$ \cite[e.g.,][]{D15}. 
In our finite simulation domain it is, however, interesting
to consider arbitrary values of $\alpha_1$.

\begin{figure}[t!]\begin{center}
\includegraphics[width=\columnwidth]{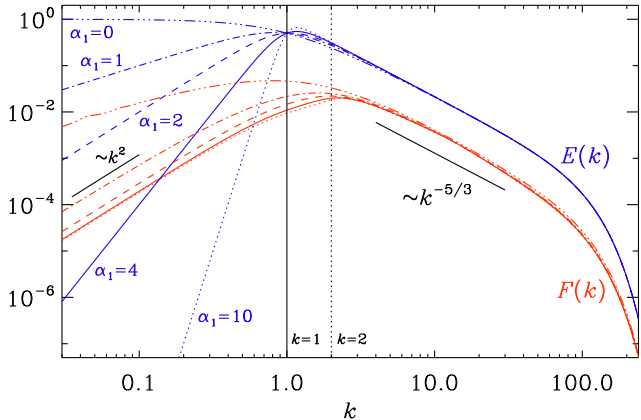}
\end{center}\caption[]{
Similar to Figure~\ref{int}, but now for the
three-dimensional case.
Blue: $E(k)$, red: $F(k)$, for
$\alpha_1=0$ (triple-dot-dashed), $1$ (dot-dashed),
$2$ (dashed), $4$ (solid), and $10$ (dotted).
}\label{int_term}\end{figure}

Figure~\ref{int_term} confirms the statement of \cite{RPMBKK19}
that $\beta=2$ is obtained
even if $E(k)$ has a blue spectrum, i.e., $\alpha\ge2$.
We also see from Figure~\ref{int_term} that for $\alpha_1=2$,
the crossover from the $k^2$ scaling for small $k$ to the $k^{-5/3}$ 
scaling for large $k$ extends now over more than one decade 
($0.2<k<5$).
This shows that we may expect slight differences when approximate
scalings are based on the inspection of spectra over a limited
dynamical range.

\begin{figure}[t!]\begin{center}
\includegraphics[width=\columnwidth]{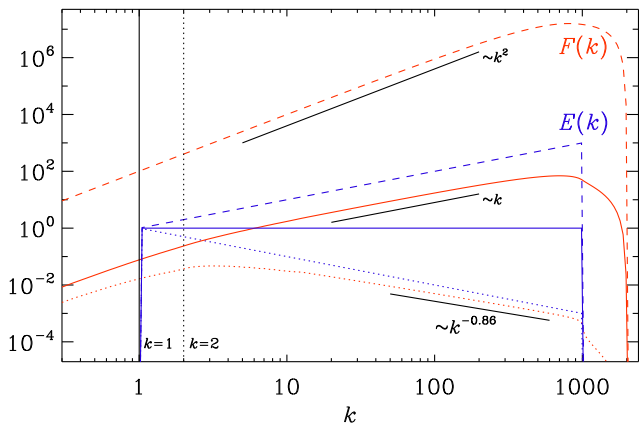}
\end{center}\caption[]{
Solutions of the 3D integral for power laws $k^\alpha$ in 
$1<k<1000$ with $\alpha=-1$, 0, and 1.
}\label{int_term3}\end{figure}

\begin{figure}[t!]\begin{center}
\includegraphics[width=\columnwidth]{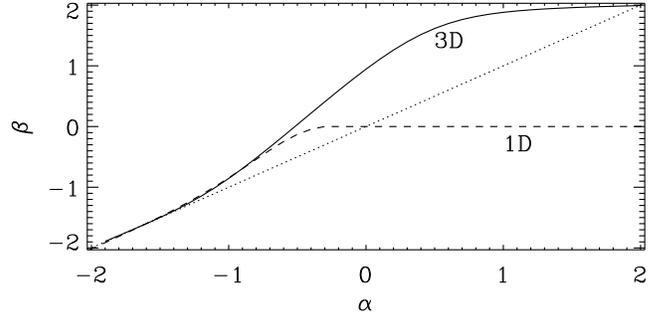}
\end{center}\caption[]{
$\beta$ versus $\alpha$ in the range from $-2$ to $2$
for one-dimensional (1D) and three-dimensional (3D) cases.
The dotted line indicates the diagonal.
}\label{pslopes}\end{figure}

To study in more detail the crossover from $\beta=2$ for
$\alpha\ge2$ to $\beta=\alpha$ for $\alpha$ of around
and below $-5/3$, for example, let us now consider
single power law spectra within a more extended
range $1\leq k\leq1000$ using
intermediate values $\alpha=-1$, 0, and 1.
No distinction between $\alpha_1$ and $\alpha_2$
will therefore be made.
The result is shown in \Fig{int_term3}.
We see that in this range of $\alpha$,
$\beta$ is always larger than $\alpha$.
We see that already for the scale-invariant $k^{-1}$ spectrum,
we have $\beta=-0.86$, so the $\beta=\alpha$ relation is only
approximately obeyed.

To determine the relation between $\alpha$ and $\beta$ in the
intermediate regime, we now compute $\beta$ using the same numerical
setup as before, but we now consider power law scalings in a range
that is 100 times larger, $1\leq k\leq 10^5$.
The result is shown in Figure~6.
Here we also compare with the corresponding results in one dimension.
We now see that in the range $-1<\alpha<1$, the value of $\beta$
deviates markedly from the $\beta=\alpha$ relation,
and that we have $\beta\approx2$ already for $\alpha=1$.

Likewise in one dimension, the $\beta=\alpha$ relation
is only true for $\alpha<-1$.
Again, there is a small intermediate interval, $-1<\alpha<0$,
where $\beta$ is somewhat larger than $\alpha$, but this departure
is by far not as dramatic as in three dimensions; see Figure~6.

\section{Comparison with turbulence simulations}

To address the assumption of Gaussianity, we now present the results
of numerical simulations of the hydromagnetic equations for a weakly
compressible gas in a cubic domain of size $L^3$ using $1024^3$ mesh
points.
We employ the {\sc Pencil Code}\footnote{\url{https://github.com/pencil-code}},
which uses sixth order accurate finite differences for the spatial
discretization and a third order time stepping scheme.
We first consider decaying nonhelical turbulence.
Our simulations are similar to Run~A of \cite{BKMRPTV17}, where
$\alpha_1\approx4$ and $\alpha_2\approx-5/3$.
The turbulence is magnetically dominated, so the velocity is almost
entirely the result of the Lorentz force.

\begin{figure}[t!]\begin{center}
\includegraphics[width=\columnwidth]{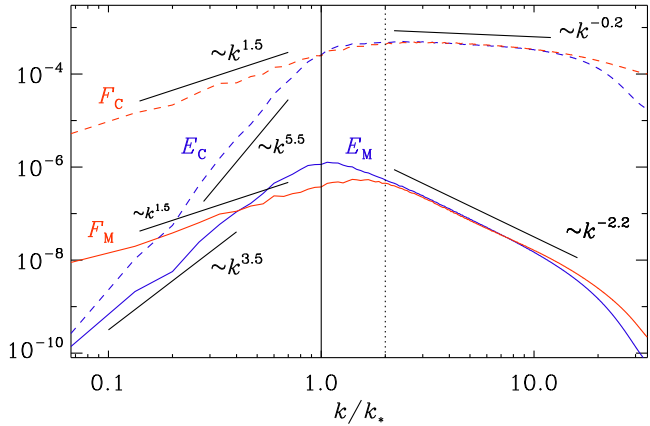}
\end{center}\caption[]{
Comparison with $E_i(k)$ and $F_i(k)$ from a turbulence simulations for the magnetic field ($i={\rm M}$) and the
current density ($i={\rm C}$).
The $F_i(k)$ spectra have been shifted to show the agreement of
their slopes with those of $E_i(k)$ for large $k$.
}\label{pstress}\end{figure}

In \Fig{pstress}, we present the results for the magnetic energy spectrum
$\EM(k)$ after about 100 Alfv\'en times.
Initially, the peak of $\EM(k)$ was at $k=k_*$ with $k_* L/2\pi=60$, but,
owing to an effect similar to the inverse cascade---here without helicity---the
peak has moved to about $k_* L/2\pi=15$ by the end of the simulation;
see \cite{BKT15} for similar results.
In \Fig{pstress} we also compare with the corresponding spectrum of
$\BB^2$, referred to as $F_i(k)$, where $i={\rm M}$ stands
for the magnetic stress.
We see that the spectral slopes of $\EM$ and $\FM$ agree in the inertial range.
In the subinertial range, the slopes of $\EM$ and $\FM$ are,
as expected, different from each other.
However, the values of the slopes, 3.5 and 1.5, respectively,
are below the expectations of 4 and 2, respectively.

To characterize the departure from Gaussianity, we have computed the
kurtosis of the magnetic field separately for all three components
and then take the average, which is denoted by
\begin{equation}
\mbox{kurt}\BB=-3+\frac{1}{3}\sum_{i=1}^3
\langle B_i^4\rangle/\langle B_i^2\rangle^2.
\end{equation}
We find a rather small value of less than 0.1.
Thus, the field is close to Gaussian and our results are qualitatively
in close agreement with those of the present paper.

To compare with a field where the assumption of Gaussianity cannot be
justified, we also show the results for the normalized current density
$\JJ=\nab\times\BB$.
We denote the corresponding spectra for current density by
$E_i(k)$ and $F_i(k)$ with $i={\rm C}$.
The kurtosis of $\JJ$, defined analogous to $\mbox{kurt}\BB$, is about 4.5.

In the inertial range, $\EM$ has a slope of about
$-2.2$, which is steeper than that of the Kolmogorov spectrum.
The current density spectrum has a slope of about $-0.2$.
The slopes were initially closer to the Kolmogorov values,
but they became steeper with time.
What is important, however, is that in the inertial range, the spectral
slopes of $\FM$ and $\FC$ agree with those of $\EM$ and $\EC$, respectively,
i.e., both have slopes of $-2.2$ and $-0.2$, respectively.
In the subinertial range, the slopes are again somewhat
different from the expectation.
We find $\alpha_1=3.5$ and $5.5$ for the spectra of $\EM$ and $\EC$,
respectively, but $1.5$ for the slopes of both $\FM$ and $\FC$.
It should be noted that, even under the assumption of isotropy, the stress
tensor contains different contributions (scalar, vector, and
tensor modes) that might behave differently.
However, the resulting differences are also sensitive to the nature
of the turbulence, whose study is beyond the scope of the present paper.

\begin{figure}[t!]\begin{center}
\includegraphics[width=\columnwidth]{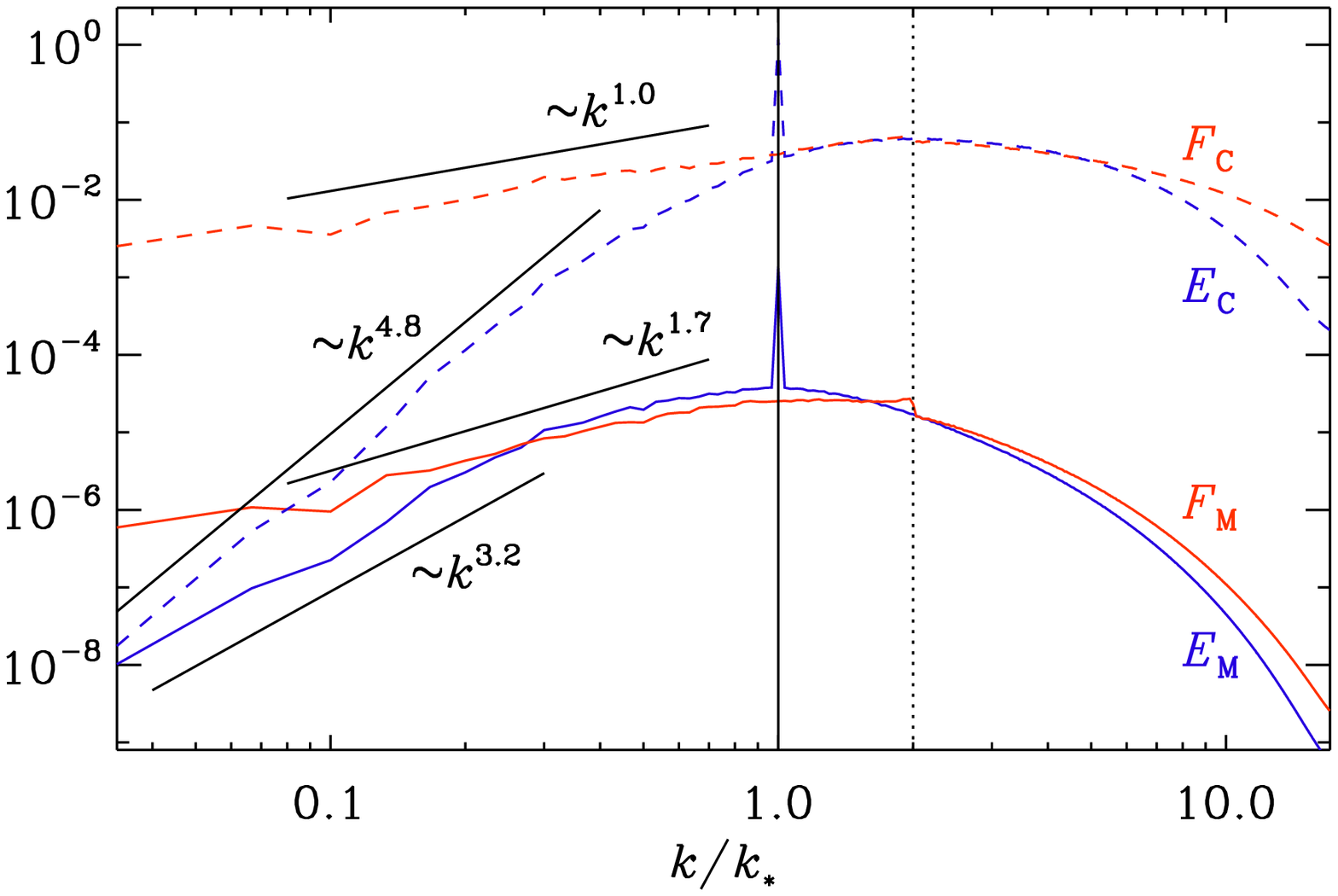}
\end{center}\caption[]{
Similar to \Fig{pstress}, but for forced turbulence
with $k_*=30$.
}\label{pstress_F1024c}\end{figure}

\begin{figure}[t!]\begin{center}
\includegraphics[width=\columnwidth]{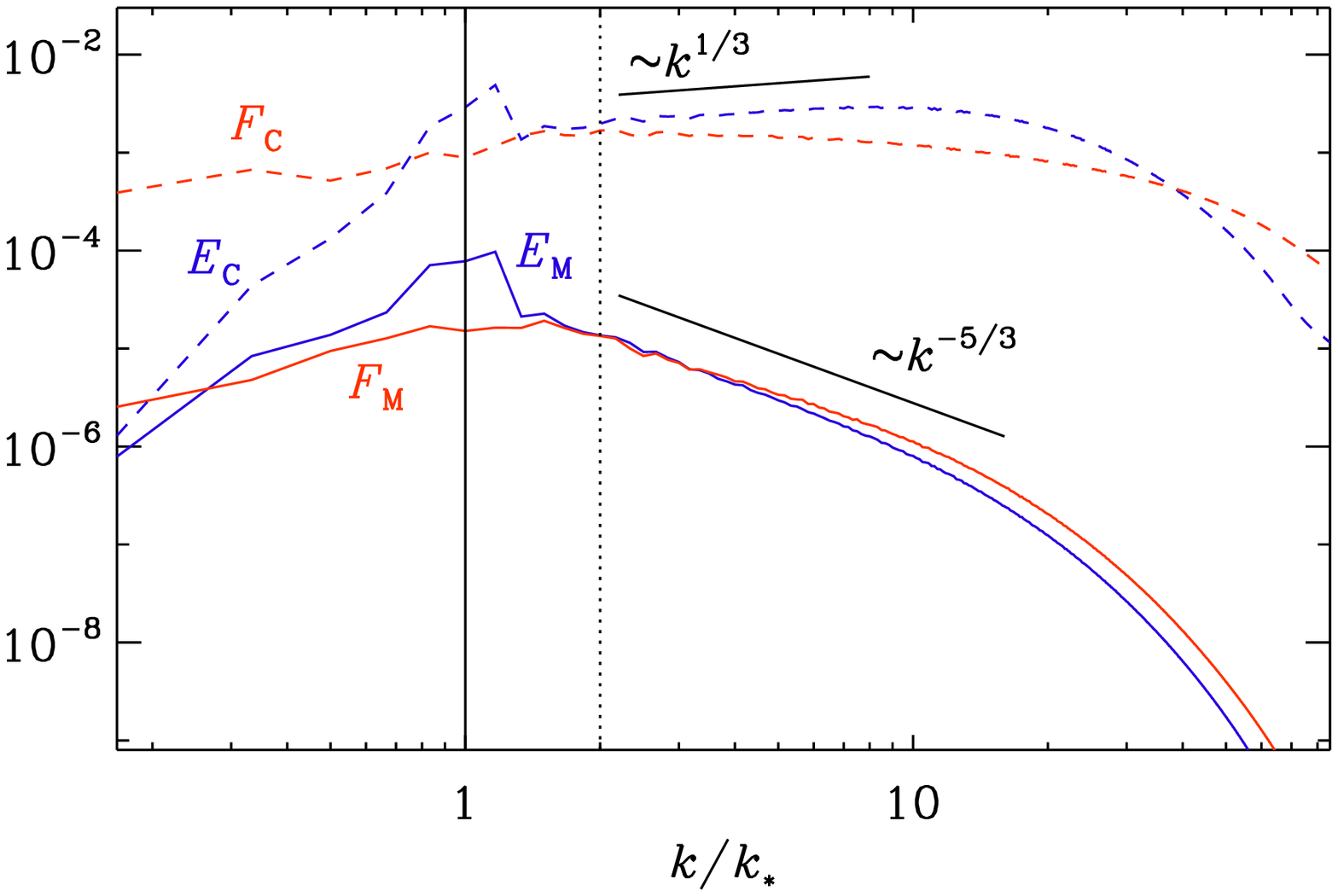}
\end{center}\caption[]{
Similar to \Fig{pstress}, but for forced turbulence
with $k_*=6$.
}\label{pstress_F1024a2}\end{figure}

Next, we compare with two runs of forced turbulence.
In these two examples, we consider the forcing wavenumbers $k_*=30$
and $6$, respectively.
In the former case (\Fig{pstress_F1024c}),
the subinertial range is more developed.
The magnetic field and current density are now closer to
being Gaussian ($|\kurt\BB|\la0.1$ and $\kurt\JJ\approx1.4$).
We see that in the subinertial range, the slope of $\FC$ is now even
more shallow ($\beta_1=1$), while that of $\FM$ is slightly steeper
($\beta_1=1.7$), but still not quite as steep as what is expected
($\beta_1=2$).

In the latter run with $k_*=6$ (\Fig{pstress_F1024a2}), the
inertial range is more developed and we see a clear
$k^{-5/3}$ spectrum in $\EM$.
The magnetic field and current density are now further away from
being Gaussian ($\kurt\BB=0.1...0.2$ and $\kurt\JJ\approx5$).
In the inertial range, the slope for $\EC$ agrees with what is
expected for Kolmogorov-type turbulence ($1/3$ for $\FM$).
However, we also see departures from the $\beta=\alpha$ relation,
where the $\beta$ for $\FM$ is slightly larger, while that of
$\FC$ is now smaller and even negative.

\section{Conclusions}

We have derived the general formula that allows us to compute a 
spectrum $F(k)$ of the square of a fluctuating field whose 
spectrum, in turn, is $E(k)$.
Our results are independent of whether the spectrum is that of
a scalar or that of a vector field.
We have seen that in the inertial range with
$E(k)\propto k^{\alpha}$ and $\alpha\la-1$, we find a spectrum
$F(k)\propto k^{\beta}$ with $\beta\approx\alpha$ if we are 
sufficiently far away from the boundaries of the validity range
of where the power law applies.
In the subinertial range, where $\alpha\ga1$, we find $\beta\approx2$.

A possible application of our work concerns the generation of
gravitational waves from hydrodynamic and hydromagnetic turbulence
with a known energy spectrum $E(k)$.
The resulting stress $T_{ij}$ sources a wave equation of the form
$\Box h_{ij}=T_{ij}$, where, except for normalization factors,
$h_{ij}$ is the linearized strain field,
$\Box=\nabla^2-c^{-2}\partial^2/\partial t^2$ is the d'Alembertian
wave operator, and $c$ is the speed of light.
The actual gravitational wave fields are the transverse and 
traceless projections of $h_{ij}$.
The nature of the wave operator can lead to a complicated relation
between the spectra of $h_{ij}$ and $T_{ij}$ \cite[e.g.,][]{RPBKKM20}.
If, however, the time delay in the wave equation can be neglected,
the spectrum of $h_{ij}$ is proportional to $F(k)/k^4$.

The simulations of \cite{RPBKKM20} show that most of the
wave generation occurs at the time when the stress has reached
maximum amplitude.
Subsequent changes of the source hardly contribute to wave production.
It may be for this reason that the assumption of no time delay
is a reasonable one.
Under this assumption, we expect that the gravitational wave energy,
which is proportional to $(\partial h_{ij}/\partial t)^2$, should be
proportional to $F(k)/k^2$; see \cite{RPBKKM20} for details.
The extent of the empirically determined departures from this
simplistic way of estimating
the gravitational wave energy spectrum are not yet fully understood 
and would need to be determined numerically or analytically,
similarly to earlier work using 
the aeroacoustic approximation of \cite{Lig52a}, as already done
by \cite{Gogo07} in the context of primordial gravitational waves.

Cosmological magnetic fields may well be helical \citep{TCFV14}.
They could be generated by the chiral magnetic effect
\citep{JS97,BFR12,Yam16,ABP17}.
However, even under the most optimistic conditions, this effect can only
produce about $10^{-18}\G$ on a scale of $1\Mpc$ at the present time
\citep{BSRKBFRK17}.
Our work now shows that the presence of magnetic helicity enhances
the stress spectrum by up to a factor of two in the subinertial range,
while leaving that in the initial range unchanged.
This implies a small shift in the peak of the stress spectrum, which
itself would hardly be a distinguishing feature.
However, helical magnetic fields lead to circular polarization of
gravitational waves \citep{KGR05}, which may be detectable with the
Laser Interferometer Space Antenna if there is a sufficiently
strong dipolar anisotropy in the signal \citep{Domcke19}.

Another potentially important application concerns the spectrum
of the parity even and parity odd linear polarization modes.
Those depend quadratically on the magnetic field components
perpendicular to the line of sight \citep{CHK17,KLP17,BBKMRPS19}.
Our work now suggests that, measuring the polarization spectrum,
one can only infer the spectrum of the underlying turbulence if
$\alpha<-1$.

\acknowledgements
\vspace{3mm}

We acknowledge initial discussions with Ethan Vishniac during
the program on the Turbulent life of Cosmic Baryons at the Aspen 
Center for Physics, which is supported by the
National Science Foundation grant PHY-1607611.
We also thank the anonymous referee for encouraging
us to study the case with helicity.
The work of AB was supported by the National Science Foundation
under the grant AAG-1615100, and
the work of SB was supported by the National Science Foundation
under the grant PHY-1707272 and by NASA under the grant
80NSSC18K0646. SB was also partly supported
by the DOE grant DE-SC0018266.
We acknowledge the allocation of computing resources provided by the
Swedish National Allocations Committee at the Center for Parallel
Computers at the Royal Institute of Technology in Stockholm.

%r e f


\begin{thebibliography}{}

\bibitem[Anand et al.(2017)]{ABP17}
Anand, S., Bhatt, J. R., \& Pandey, A. K.\yjcap{2017}{07}{051}

\bibitem[Boyarsky et al.(2012)]{BFR12}
Boyarsky, A., Fr\"ohlich, J., \& Ruchayskiy, O.\yprl{2012}{108}{031301}

\bibitem[Brandenburg et al.(2019)]{BBKMRPS19}
Brandenburg, A., Bracco, A., Kahniashvili, T., Mandal, S., Roper Pol, A., Petrie, G. J. D., \& Singh, N. K.\yapj{2019}{870}{87}

\bibitem[Brandenburg et al.(2017)]{BKMRPTV17}
Brandenburg, A., Kahniashvili, T., Mandal, S., Roper Pol, A., Tevzadze, A. G., \& Vachaspati, T.\yprd{2017}{96}{123528}

\bibitem[Brandenburg et al.(2015)]{BKT15}
Brandenburg, A., Kahniashvili, T., \& Tevzadze, A. G.\yprl{2015}{114}{075001}

\bibitem[Brandenburg et al.(2017)]{BSRKBFRK17}
Brandenburg, A., Schober, J., Rogachevskii, I., Kahniashvili, T., Boyarsky, A., Fr\"ohlich, J., Ruchayskiy, O., \& Kleeorin, N.\yapjl{2017}{845}{L21}

\bibitem[Caldwell et al.(2017)]{CHK17}
Caldwell, R. R., Hirata, C., \& Kamionkowski, M.\yapj{2017}{839}{91}

\bibitem[Davidson(2015)]{D15}
Davidson, P. A.\ybook{2015}{Turbulence: An introduction for scientists and engineers}{Oxford: Oxford University Press}

\bibitem[Domcke et al.(2019)]{Domcke19}
Domcke, V., Garcia-Bellido, J., Peloso, M., Pieroni, M., Ricciardone, A., Sorbo, L., \& Tasinato, G.\arxiv{2019}{1910.08052}

\bibitem[Durrer et al.(2000)]{DFK00}
Durrer, R., Ferreira, P. G., \& Kahniashvili, T.\yprd{2000}{61}{043001}

\bibitem[Durrer \& Caprini(2003)]{DC03}
Durrer, R., and Caprini, C.\yjour{2003}{J. Cosmol. Astropart. Phys.}{0311}{010}

\bibitem[Gogoberidze et al.(2007)]{Gogo07}
Gogoberidze, G., Kahniashvili, T. and Kosowsky, A.\yprd{2007}{76}{083002}

\bibitem[Joyce \& Shaposhnikov(1997)]{JS97}
Joyce, M., \& Shaposhnikov, M.\yprl{1997}{79}{1193}

\bibitem[Kahniashvili et al.(2005)]{KGR05}
Kahniashvili, T., Gogoberidze, G., \& Ratra, B.\yprl{2005}{95}{151301}

\bibitem[Kamionkowski et al.(1994)]{KKT94}
Kamionkowski, M., Kosowsky, A., \& Turner, M.\yprd{1994}{49}{2837}

\bibitem[Kandel et al.(2017)]{KLP17}
Kandel, D., Lazarian, A., \& Pogosyan, D.\ymn{2017}{472}{L10}

\bibitem[Lighthill(1952a)]{Lig52a}
Lighthill, M. J.\yjour{1952a}{Proc. Roy. Soc. Lond. A}{211}{564}

\bibitem[Lighthill(1952b)]{Lig52b}
Lighthill, M. J.\yjour{1952b}{Proc. Roy. Soc. Lond. A}{222}{1}

\bibitem[Lithwick \& Goldreich(2001)]{LG01}
Lithwick, Y., \& Goldreich, P.\yapj{2001}{562}{279}

\bibitem[Roper Pol et al.(2020)]{RPBKKM20}
Roper Pol, A., Brandenburg, A., Kahniashvili, T., Kosowsky, A., \& Mandal, S.\ygafd{2020}{114}{130}

\bibitem[Roper Pol et al.(2019)]{RPMBKK19}
Roper Pol, A., Mandal, S., Brandenburg, A., Kahniashvili, T., \& Kosowsky, A.\sprd{2019}
{1903.08585}

\bibitem[Stein(1967)]{Ste67}
Stein, R. F.\ysph{1967}{2}{385}

\bibitem[Subramanian(2016)]{Sub16}
Subramanian, K.\yrpp{2016}{79}{076901}

\bibitem[Tashiro et al.(2014)]{TCFV14}
Tashiro, H., Chen, W., Ferrer, F., \& Vachaspati, T.\ymn{2014}{445}{L41}

\bibitem[Yamamoto(2016)]{Yam16}
Yamamoto, N.\yprd{2016}{93}{125016}

\end{thebibliography}
\end{document}